\documentclass{aa}
\usepackage{lscape,color,graphicx}
\usepackage[varg]{txfonts}

\begin{document}
   \title{The effect of ambipolar resistivity on the 
	formation of dense cores}
   \author{S. Van Loo\inst{1}, S. A. E. G. Falle\inst{2},
        T. W. Hartquist\inst{1} \and A. J. Barker\inst{1,3}}
   \authorrunning{Van Loo et al.}

   \offprints{S. Van Loo\\ \email{svenvl@ast.leeds.ac.uk}}

   \institute{School of Physics and Astronomy, University of Leeds,
        Leeds LS2 9JT, UK \and School of Mathematics, University of Leeds,
        Leeds LS2 9JT, UK \and Department of Applied Mathematics and 
	Theoretical Physics, University of Cambridge, Cambridge CB3 0WA, UK} 

   \date{Received date; accepted date}

   \abstract
        {}
        {We aim to understand the formation of dense cores by magnetosonic 
	waves in regions where the thermal to magnetic pressure ratio is small.
        Because of the low-ionisation fraction in molecular clouds, neutral
	and charged particles are weakly coupled. Ambipolar
	diffusion then plays an important r\^ole in the formation process.}
        {A quiescent, uniform plasma is perturbed by a fast-mode 
         wave. Using 2D numerical simulations, we follow the evolution of 
	the fast-mode wave. The simulations are done with a multifluid,
	adaptive mesh refinement MHD code.}
        {Initial perturbations with wavelengths that are 2 orders of magnitude 
	larger than the dissipation length are strongly affected by the 
        ion-neutral drift. Only in situations where there are large 
        variations in the magnetic field corresponding to a highly 
	turbulent gas can fast-mode waves generate dense cores. This means 
	that, in most cores, no substructure can be produced. 
	However, Core D of TMC-1 is an exception to this case. Due to 
	its atypically high ionisation fraction, waves with wavelengths up 
	to 3 orders of magnitude greater than the dissipation length
	can be present. Such waves are only weakly affected 
        by ambipolar diffusion and can produce dense substructure without 
	large wave-amplitudes.
        Our results also explain the observed transition from 
	Alfv\'enic turbulent motion at large scales to subsonic motions
  	at the level of dense cores.}
        {}


   \maketitle

\section{Introduction}
Observations show that molecular clouds are highly structured
(e.g. Blitz \& Stark \cite{BS85}). Furthermore, emission-line profiles of  
molecular tracers such as CO and CS are considerably broader than their 
thermal line widths indicating the presence of highly turbulent motions
(e.g. Falgarone \& Phillips \cite{FP90}). Since molecular clouds are 
threaded by magnetic fields, it is natural to suppose that 
these observed line widths are due to magnetohydrodynamic (MHD) waves 
(Arons \& Max \cite{AM75}). 

Large-scale, three-dimensional (3D)
simulations of turbulent gas motions (e.g. Ballesteros-Paredes \& 
Mac Low \cite{BM02}; Padoan \& Nordlund \cite{PN02};
Gammie et al.~\cite{Getal03}; Li et al.~\cite{Letal04}; 
Galv{\'a}n-Madrid et al.~\cite{Getal07}) show that dense cores with
statistical properties similar to those observed can
indeed be formed in this way.
By following the evolution of a single MHD wave, Falle \& Hartquist 
(\cite{FalleHartquist02}) in 1D and Van Loo et al.~(\cite{VanLooetal06}) 
in 2D found that dense cores could be generated by the excitation of 
slow-mode waves. This process works on different length-scales 
and can also explain the formation of substructure within cores themself
(Van Loo et al.~\cite{VanLooetal07}).

In all these simulations, the plasma is treated as a single ideal plasma
in which the plasma is perfectly coupled with the magnetic field. 
However, the low ionisation fraction in molecular clouds (Elmegreen
\cite{E79}) implies that
the plasma and magnetic field are actually weakly coupled on the scale
of the cores. The charged 
particles then drift through the neutral particles giving rise to 
ambipolar diffusion (Mestel \& Spitzer \cite{MS56}). Large-scale, 3D
simulations including ambipolar resistivity (Padoan et al.~\cite{PZN00};
Oishi \& Mac Low \cite{OM06}) show that dense cores still arise in those
conditions.
Lim et al.~(\cite{Limetal05}) investigated in 1D the effect of ambipolar 
resistivity on the evolution of a single MHD wave. They found that 
ambipolar diffusion affects the evolution of waves with wavelengths
up to about a thousand times the dissipation length-scale and thus
has a significant effect on much of the observed structure in star 
formating regions. 

In this paper we extend the model of Lim et al.~(\cite{Limetal05})
to two dimensions. The governing equations and the initial condition are 
given in Sect.~\ref{sect:model}. We then examine 
the density structure in the numerical calculations 
(Sect.~\ref{sect:results}).
and discuss the relevance of these results in Sect.~\ref{sect:discuss}.

\section{The model}\label{sect:model}
\subsection{Governing equations}
Since the ionisation fraction within molecular clouds is low, the plasma
needs to be treated as a multicomponent fluid consisting of neutrals,
and charged particles. Here we assume the charged particles to be 
 ions and electrons only. In the limit of small mass 
densities for the charged fluids, their inertia can be neglected. 
The governing equations for the neutral fluid are given by
\begin{eqnarray}\label{eq:neutral}
    \frac{\partial{\rho_n}}{\partial{t}} + \nabla (\rho_n {\bf v}_n)  =  0,\\
    \frac{\partial{\rho_n {\bf v}_n}}{\partial{t}} + \nabla  (\rho_n {\bf v}_n
        {\bf v}_n + p_n)  =  {\bf J}\times{\bf B}, \nonumber
\end{eqnarray}
with $\nabla = (\partial/\partial{x}, \partial/\partial{y})$. For
the charged fluid $j$, the equations reduce to 
\begin{eqnarray}\label{eq:charged}
     \frac{\partial{\rho_{j}}}{\partial{t}} + \nabla (\rho_{j} 
        {\bf v}_{j})  =  s_{jn},\\    
     \alpha_{j} \rho_{j} ({\bf E} + {\bf v}_j\times {\bf B}) + 
        \rho_{j} \rho_n K_{jn} ({\bf v}_n - {\bf v}_j)  =  0. \nonumber
\end{eqnarray}
where $\alpha_j$ is the charge to mass ratio, $K_{jn}$ the collision
coefficient of the charged fluid $j$ with the neutrals, $s_{jn}$ 
the mass transfer rate between the charged fluid $j$ and the neutral 
fluid, $\bf{B}$ the magnetic field, $\bf{E}$ the electric field and 
$\bf{J}$ the current given by $\bf{J} = \sum_j{\alpha_j \rho_j \bf{v}_j}$.  
We adopt an isothermal equation of state $p = \rho T$ for all fluids
with $T$ the isothermal temperature. Note that the units 
are chosen so that all factors of $4\pi$ and $c$ disappear
from the equations.

Within translucent clumps the ionisation fraction, $X_i$, changes as a function 
of the neutral density (e.g. Ruffle et al.~\cite{Ruffleetal98}). For a visual 
extinction of $A_V = 3$, the ionisation fraction decreases 
from  $X_i = 10^{-4}$ for $n_H \leq 1000~{\rm cm^{-3}}$ to 
$X_i = 10^{-7}$ for $n_H = 5\times10^4$~cm$^{-3}$, roughly following 
a $1/n_H$ law. As the ionisation and recombination rates are sufficiently
high in molecular clouds, ionisation equilibrium is quickly reached.
We therefore adopt mass transfer rates $s_{jn}$ so that the fractional 
abundances of ions and electrons vary as $1/n_H$.
 
The Hall parameters of the ions and electrons, i.e. the ratio between
their gyrofrequency and the collision frequency with the neutrals, 
are large in most astrophysical situations
(Wardle~\cite{Wardle98}). The ions and electrons thus gyrate many times
around a magnetic field line before they collide with a neutral particle.
These collisions affect the evolution of the magnetic field which is  
governed by     
\begin{equation}\label{eq:Bfield}
	 \frac{\partial{{\bf B}}}{\partial{t}} - \nabla \times       
        ({\bf v}_n \times {\bf B}) = \nabla \times \left(r_a 
	\frac{((\nabla \times {\bf B})\times{\bf B})\times{\bf B}}{B^2}\right),
\end{equation}
where $r_a$ is the ambipolar resistivity (Falle~\cite{Falle03}).
As the electrons have a Hall parameter which is larger than 
the ions, the ambipolar resistivity is given by
\[
	r_a = \frac{B^2}{\rho_i \rho_n K_{in}}.
\]

Equations~\ref{eq:neutral} -~\ref{eq:Bfield} are solved with an 
adaptive mesh refinement code using the scheme described by 
Falle~(\cite{Falle03}) which we extended to two dimensions. 
This scheme uses a second-order Godunov solver
for the neutral fluid equations (Eqs.~\ref{eq:neutral}a and b). 
The charged fluid densities are calculated using an explicit 
upwind approximation to the mass conservation equation 
(Eq.~\ref{eq:charged}a), while the velocities can be calculated 
from the reduced momentum equations (Eq.~\ref{eq:charged}b). 
The magnetic field is advanced explicitly, even though this implies a 
restriction on  the stable time step at high numerical resolution due to the 
ambipolar resistivity term, i.e. $\Delta t < \Delta x^2/4r_a$ 
(Falle \cite{Falle03}). 

The code uses a hierarchy of grids such that the grid 
spacing of level n is $\Delta x/2^n$, where $\Delta x$ the grid spacing 
of the coarsest level. The solution is computed on all grids and the
difference between the solutions on neighbouring levels is used to
control refinement. 
More specifically, if the mapped down coarse cell value of the neutral flow 
variables differs by more than 10\% from the fine cell value, the grid is 
refined.
The loss in efficiency 
due to the time step restriction is then partially balanced by the gain due 
to adaptive mesh refinement. We also implemented the divergence cleaning
algorithm of Dedner et al.~(\cite{Dedneretal02}) to eliminate the errors
due to non-zero $\nabla \cdot {\bf B}$.

\subsection{Initial conditions}\label{sect:initial}
Like Lim et al.~(\cite{Limetal05}) we study the formation 
of dense inhomogeneities by following the evolution of a fast-mode
wave in a uniform plasma in which the magnetic pressure dominates over
the gas pressure. 
Slow-mode waves are not excited in the initial state. They have a very 
low propagation speed, and will therefore remain localised near the
boundary until they are generated inside the cloud by  nonlinear 
steepening of the fast-mode wave.

The quiescent, uniform  background plasma conditions are given 
(in dimensionless units) by
\[
	\rho_n = 1, \rho_e = 5\times10^{-8}, \rho_i = 10^{-3}, {\bf v}_{n,i,e} 
	= 0, B_x = 1, B_y = 0.25, 
\]
and an isothermal temperature $T = 0.015$ for all fluids. Furthermore,
we adopt $\alpha_e = -2.0\times 10^{12}$ and $\alpha_i = 10^{8}$ for the 
electron and ion charge to mass ratio, and $K_{en} = 4.0\times10^5$ 
and  $K_{in} = 2.0\times10^4$ for the electron and ion collision coefficients. 
These values correspond to the properties of a translucent clump with optical 
extinction $A_V = 3$, a neutral number density of 500~cm$^{-3}$ 
and a magnetic field of 30~$\mu$G for a core temperature of 10~K. 
Note that we adopt $m_n = 2 m_H$ and $m_i = 10 m_H$ for the neutral and 
ion mass. The adopted value for the ion mass lies well within the observed 
range, i.e. from $\approx 3 m_H$ when H$^+_3$ is the dominant ion to 
$\approx 30 m_H$ when it is HCO$^+$ (Caselli et al.~\cite{Casellietal02}).
  
A non-linear fast magnetosonic wave propagating in the positive 
$x$-direction is then superposed onto the 
uniform background. The initial state needs to be calculated using the 
method described in Lim et al.~(\cite{Limetal05}) and Van Loo et 
al.~(\cite{VanLooetal07}), since a simple linear approximation is not 
valid. This method requires that a wave satisfies  
\[
	\frac{\partial{{\bf P}}}{\partial{x}} \propto {\bf r}_f,
\]
where ${\bf P}$ is the vector of the primitive variables $\rho,
{\bf v}$ and ${\bf B}$, and ${\bf r}_f$
the right fast-mode eigenvector (see Falle \& Hartquist 
\cite{FalleHartquist02}).  By specifying the profile for $v_y(x,y)$,
all other primitive variables can be easily calculated. We assume that, 
for all fluids, the $y$-component of the velocity associated with the 
wave changes sinusoidally with $x$.
Furthermore, we introduce a sinusoidal phase-shift with respect to the
$y$-direction to make the flow two-dimensional, so that 
\[
	v_y(x,y) = A_1 \sin\left[\frac{2\pi}{\lambda}\left(x + A_2
	\sin\left(\frac{2\pi y}{\lambda_2}\right) \right) \right], 
\]
where $\lambda$ is the wavelength of the fast-mode wave, 
$\lambda_2$ the wavelength of the perturbation in the $y$-direction and
$A_2$ the amplitude of the phase shift in the $y$-direction. 
$A_1$ is the amplitude of the $y$-component of the velocity and 
is chosen so that the maximum amplitude of the total velocity 
in the wave ($\sqrt{v^2_x + v^2_y})$ is a specified value $A$. 
We perform calculations for $A = 0.7$, 1.0 and 1.5. As the Alfv\'en
speed, $v_a = B/\sqrt{\rho}$, is roughly 1.0 in our simulations, these 
models correspond to sub-, trans- and super-Alfv\'enic
perturbations. As in
to Van Loo et al.~(\cite{VanLooetal06}), we set $A_2 = 0.01 \lambda$ 
and $\lambda_2 = \lambda$. The spatial variations in the 
fast-mode waves  
with different wavelength $\lambda$ are then identical and any differences 
in the dynamical evolution is due to the physical processes and not the 
initial conditions.  

Due to the ambipolar resistivity in a weakly  ionised plasma, a fast-mode 
magnetosonic wave of long wavelength $\lambda$ dissipates on a time-scale  
\[
	t_d = \frac{\lambda^2}{r_a}.
\]
Since we are interested in the generation of dense structures associated 
with slow-mode waves, it is necessary that the initial fast-mode wave
can generate these slow-mode waves before it dissipates. The 
time for the wave to steepen into a shock is roughly
\begin{equation}\label{eq:ts}
	t_s = \frac{\lambda}{2A_1},
\end{equation}
which gives the following condition
\[
	\lambda >> \frac{r_a}{2A_1} = \frac{0.025}{A_1},
\]
for dissipation to be unimportant initially. Like Lim et 
al.~(\cite{Limetal05}), we study the evolution of fast-mode waves 
with $\lambda = \pi$ and $10\pi$ which satisfy the above condition. 
The numerical domain is 
$0 \le x \le \lambda$ and   $0 \le y \le \lambda$ with periodic
boundary conditions. We use 4 grid levels with a finest 
grid resolution $800 \times 800$. 

We also do simulations for $\lambda = 100\pi$, but these are done 
with the ideal MHD code described in Falle (\cite{Falle91}). 
This is because  we cannot resolve shocks if $\lambda = 100~\pi$ with 
the above resolution. The 
physical 
shock thickness, which is of the order of the ambipolar 
dissipation length $l_d$, is smaller than the finest grid spacing.
(The dissipation length $l_d$ is the wavelength of the wave that 
dissipates in one wave period, i.e. $c_f t_d = \lambda$ with $c_f$ 
the fast-magnetosonic speed and $l_d = 0.05$ in our simulations.)
This itself is not a problem as the numerical scheme 
smears out the shock front over a few grid cells. However,  
the magnetic field then appears discontinuous on the grid and  
this cannot be handled by our numerical procedure (see 
Falle \cite{Falle03}). 
Although we
could use higher numerical resolution this increases the computational
cost considerably and is unnecessary since the results of Lim et
al.~(\cite{Limetal05}) show that dissipation is negligible for 
$\lambda \ge 30\pi$  and the flow can 
be treated as a single ideal fluid.

\section{Results}\label{sect:results}
\subsection{Large wavelengths, e.g. $\lambda = 100\pi$}
It is useful to first consider the evolution of fast-mode waves 
in the absence of ambipolar dissipation. This is not only an 
appropriate description for waves with wavelengths larger than 
$\approx 30 \pi$ (see above), it also allows us to properly describe 
the effect of ambipolar dissipation.

\begin{figure}
\resizebox{\hsize}{!}{\includegraphics{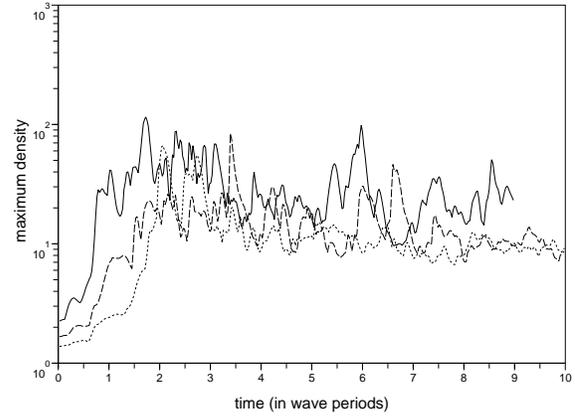}}
\caption{Temporal evolution of the maximum 
	normalised number density for $\lambda = 100\pi$ with 
   	different wave amplitudes: A = 1.5 (solid), 1.0 (dashed) and
	0.7 (dotted).
	The normalisation factor is $n_H = 10^3$~cm$^{-3}$.
	}
\label{fig:ideal}
\end{figure}

Van Loo et al.~(\cite{VanLooetal06}) performed similar simulations in which
they followed small-amplitude fast-mode waves.
These simulations show that 
the first dense structures are formed by the non-linear steepening of
the wave and that the subsequent interaction of the fast-mode shock
with these dense regions generates additional structure. 
Since non-linear steeping occurs earlier for larger initial amplitudes
(see Eq.~\ref{eq:ts}), we would expect dense structures to appear 
earlier for such waves. Furthermore, it was shown that larger wave amplitudes
result in larger density contrasts, i.e. $\rho_{\rm max} \propto A_1^2$.
Figure~\ref{fig:ideal} shows that this dependence still holds 
for finite-amplitude waves.

\begin{figure}
\resizebox{\hsize}{!}{\includegraphics{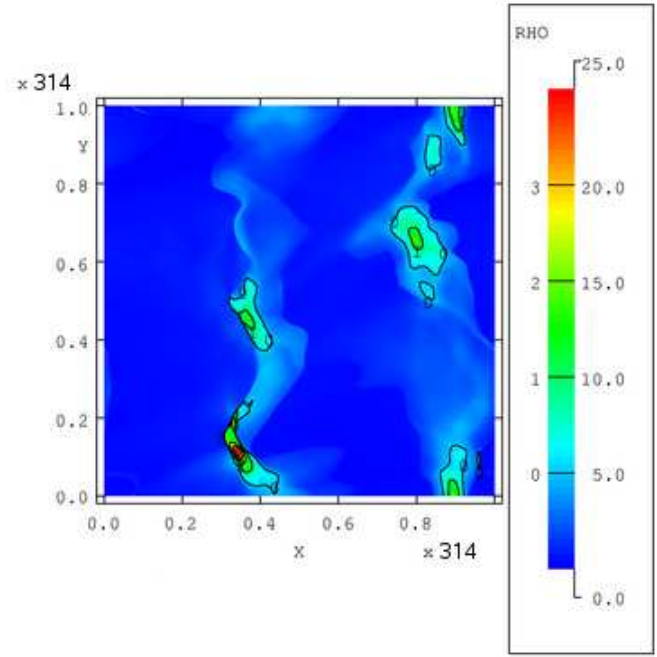}}
\caption{Density structure for $\lambda = 100 \pi$ with wave amplitude 
	$A_1 = 1.5$ after 4 wave periods. The contour lines indicate
 	normalised number densities higher than 5, 10, 15 and 20. The 
	same normalisation factor as in Fig.~\ref{fig:ideal} is used.
        }
\label{fig:rho_struct}
\end{figure}

The density structure for finite-amplitude waves is more filamentary than 
produced by the small-amplitude waves. This is a result of the more 
rapid steepening of the fast-mode wave and the concurrent faster growth of the 
perturbation in the $y$-direction. Figure~\ref{fig:rho_struct} shows that
some substructure with $\rho > 10 $, or $n_H > 10^4$~cm$^{-3}$
($\rho = 1$ corresponds to $n_H = 10^3$~cm$^{-3}$, see 
Sect.~\ref{sect:initial}), is present within these filaments. As these 
density structures have densities higher than the threshold for line 
emission of NH$_3$, they can, in a traditional sense, be referred to as 
{\it dense cores} (e.g. di Francesco et al.~\cite{dFetal07}). 
In our simulations,
dense cores arise where a 
shock interacts with another shock or with an already present dense region.  
In such interactions slow-mode waves are excited which are
associated with density enhancements 
(e.g. Falle \& Hartquist \cite{FalleHartquist02}; Van Loo et 
al.~\cite{VanLooetal06}). 

Although the initial fast-mode wave is not affected by ambipolar dissipation,
it can be expected that the dense cores are. However, 
slow-mode waves associated with dense cores do not generate large 
perturbations in the magnetic field (see e.g. Falle \& Hartquist 
\cite{FalleHartquist02}). 
(Note that this means that the magnetic field within a dense core 
is roughly the same as the background magnetic field.)
As ambipolar diffusion only acts on the magnetic field, slow-mode waves 
propagate without any significant damping
(Balsara~\cite{Balsara96}).
The lifetimes 
of gravitationally unbound dense cores are, therefore, not determined by 
the ambipolar dissipation, but
by the dispersion time-scale of a core.  This dispersion time scale is 
given by $t_{\rm disp} \approx r/v_a$, where $r$ is the core radius,
and is generally short compared to the wave 
period of the initial wave.  Figure~\ref{fig:ideal} shows that high-density 
contrasts are present for several wave periods. 
Slow-mode excitation and the associated dense core formation only
ceases when the fast-mode shock has dissipated away most of its energy.
The formation of structure is therefore driven by the fast-mode wave.    

\subsection{Intermediate wavelengths, e.g. $\lambda = 10\pi$}
\begin{figure}
\resizebox{\hsize}{!}{\includegraphics{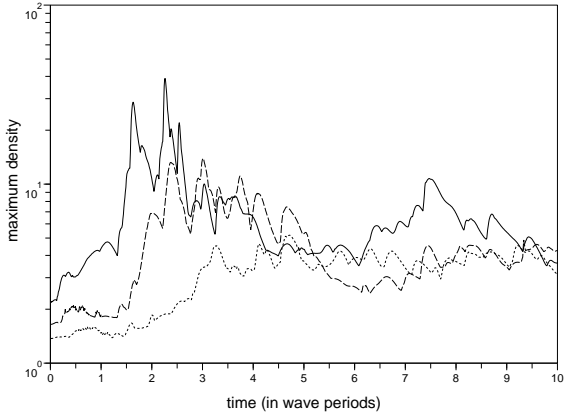}}
\caption{Same as Fig.~\ref{fig:ideal} but for $\lambda = 10\pi$.}
\label{fig:mid}
\end{figure}

For the wave-amplitudes studied here, a fast-mode wave with a 
wavelength $\lambda = 10 \pi$ steepens into a shock before ambipolar 
diffusion becomes important.  Although this suggests 
a similar evolution for waves with $\lambda = 10\pi$ as for 
$\lambda =100\pi$, there are some important differences.
Figure~\ref{fig:mid} shows that the high-density structures develop
later than for $\lambda = 100\pi$ (see Fig.~\ref{fig:ideal}). Also,
the maximum densities within cores are roughly a factor of 5-10 smaller.
Note that for $\lambda = 10~\pi$, $\lambda/l_d = 200~\pi$ and that ambipolar
resistivity is thus important even for waves with wavelengths that are 
rather large compared to the dissipation length.

\begin{figure}
\resizebox{\hsize}{!}{\includegraphics{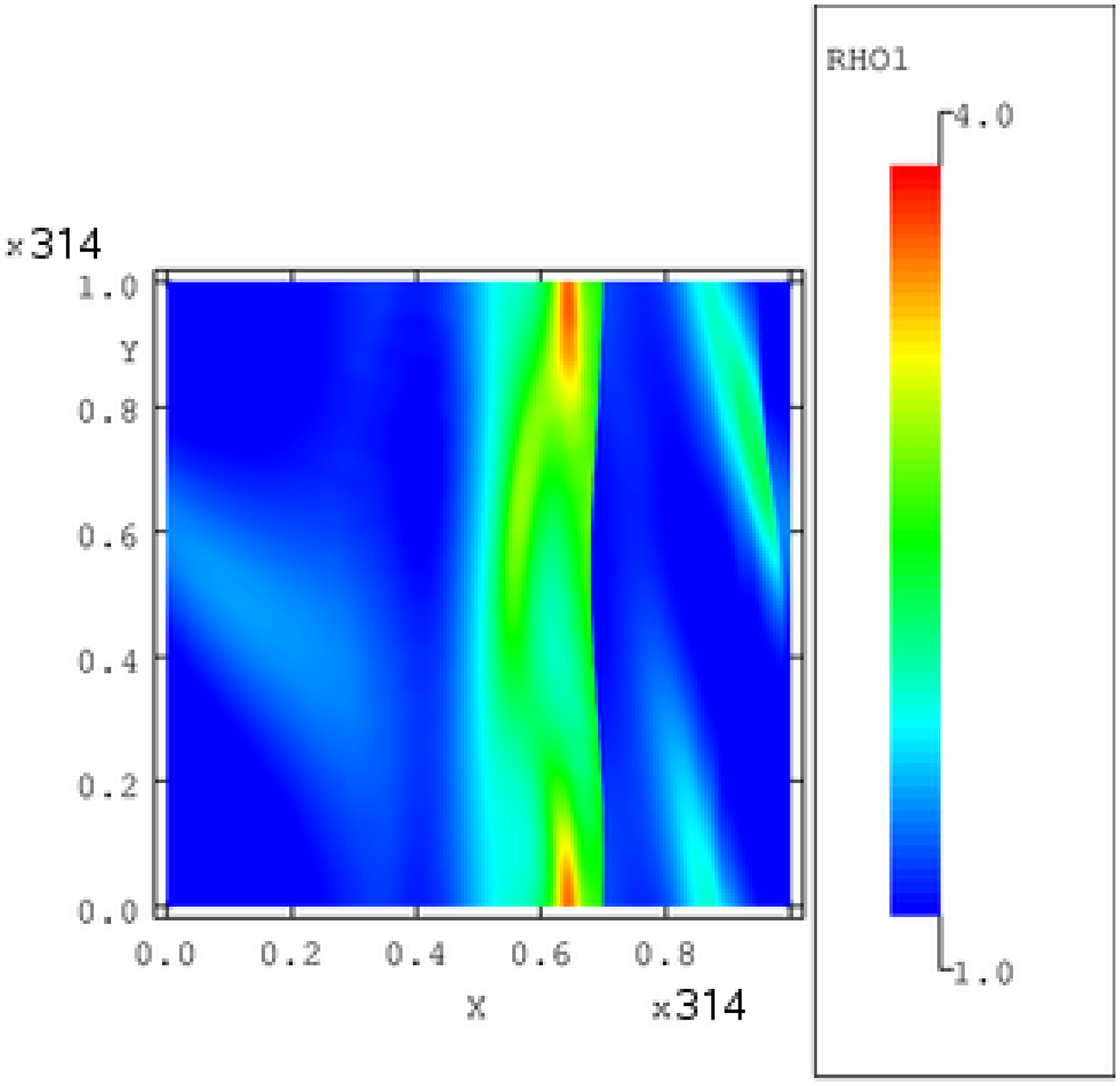}}
\resizebox{\hsize}{!}{\includegraphics{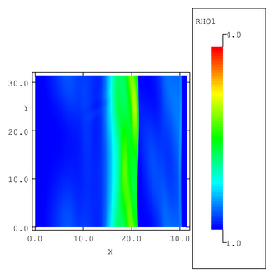}}
\caption{The normalised 
	density structure at half a wave period for $\lambda = 100\pi$
	(upper panel) and $\lambda = 10\pi$ (lower panel) with A=1.5. 
	The density range is set between 1 and 4 
	with the same normalisation factor as in Fig.~\ref{fig:ideal}.}
\label{fig:diff}
\end{figure}

It is not the fast-mode wave that is significantly affected by ambipolar
diffusion, but the initial perturbation perpendicular to the propagation 
direction is. So far, we have neglected its r\^ole in the dynamics. 
The initial $y$-dependent contribution of the perturbation due to the 
phase-shift of the fast-mode wave  
has the same wavelength as the fast-mode wave, but the amplitude 
of the variations in the magnetic field due to this phase shift 
are about two orders of magnitude (or more) smaller. The variations in
the $y$-direction will therefore dissipate more quickly than 
the those in the $x$-direction. Evidently, the variations perpendicular 
to the magnetic field grow more slowly. Figures~\ref{fig:diff} 
a and b show the difference in the growth of these variations for 
$\lambda = 100\pi$ and $\lambda = 10\pi$.

As a result, it becomes more difficult to produce large density 
perturbations. The largest density 
inhomogeneities form within dense filaments due to the interaction of 
these regions with shocks (e.g.~Van Loo et al.~\cite{VanLooetal06}). 
The core density  thus depends on the density of the parent 
filament and of the shock strength. In this case, not only do the
filaments have smaller densities than for $\lambda = 100\pi$, but the shock strength is continuously decreasing due to the ambipolar 
dissipation. This also explains why there is only a short period of time
(i.e.  $\approx$ 3-4 wave periods) for which we find peak densities 
with $\rho > 10$ 
or $n_H > 10^4$~cm$^{-3}$.

\subsection{Short wavelengths, e.g. $\lambda = \pi$}
\begin{figure}
\resizebox{\hsize}{!}{\includegraphics{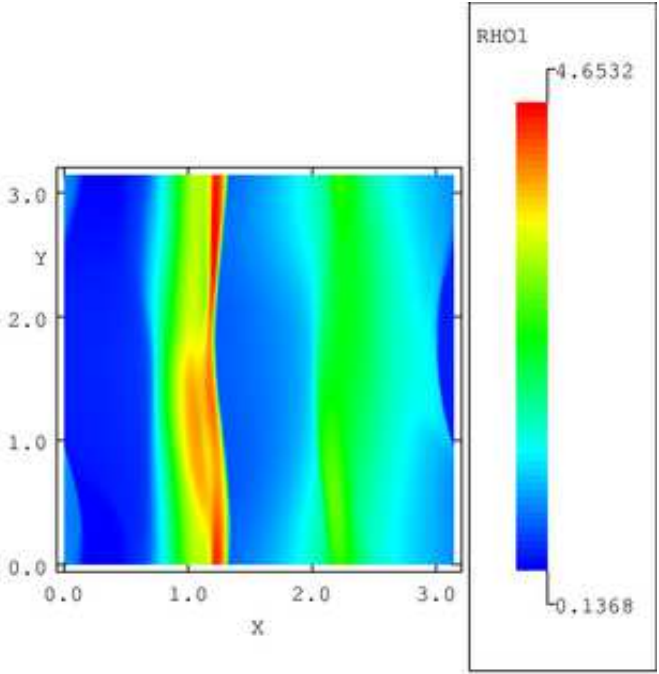}}
\caption{The 
	normalised
	density structure after 3 wave periods for $\lambda = \pi$ 
with $A=1.5$.}
\label{fig:shortt}
\end{figure}

In the previous section, we found that ambipolar dissipation has a
moderate effect on 
the evolution of waves with a wavelength $\lambda = 10\pi$. For shorter
wavelengths, i.e. $\lambda = \pi$, the impact is greater even though 
the wave is two orders of magnitude longer than the dissipation length.
The initial $y$-dependent contribution to the perturbation 
is, as can be expected from 
the previous case, not large enough to develop any significant
variation in that direction. This means that there are only variations in the 
$x$-direction (see Fig.~\ref{fig:shortt}). The evolution of the wave is 
then nearly one-dimensional. 

\begin{figure}
\resizebox{\hsize}{!}{\includegraphics{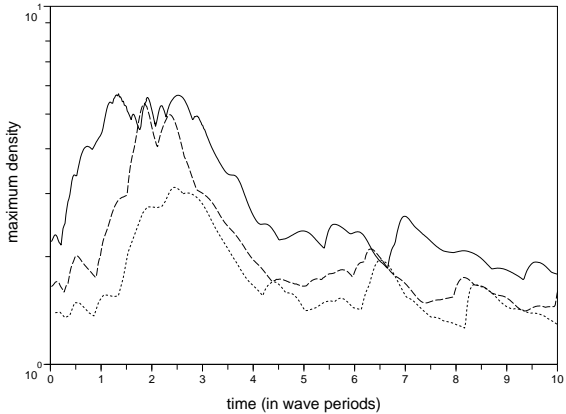}}
\caption{Same as Fig.~\ref{fig:ideal} but for $\lambda = \pi$.}
\label{fig:short}
\end{figure}

Ambipolar dissipation causes rapid decay of the fast-mode wave:
the shock wave loses its energy within a few wave periods. The
generation of density inhomogeneities then stops, 
as can be seen in Fig.~\ref{fig:short}. The dissipation process, however,
also contributes  to the formation of density structures at short wavelengths.
Lim et al.~(\cite{Limetal05}) showed that, for $\lambda \approx \pi$,
ambipolar dissipation plays a major r\^ole in the excitation of slow-mode
waves. Figure~\ref{fig:short} indeed shows that the density contrasts 
arise earlier than for $\lambda = 10\pi$. However, ambipolar diffusion
is still more efficient in reducing the peak density which does not 
become much higher than $\rho = 5$ 
or $n_H = 5\times 10^3$~cm$^{-3}$. 

To break the one dimensionality of the flow and form cores, the initial
phase shift needs to be larger than in our standard model: the variations of 
the magnetic field in the $y$-direction should be of the same order 
as the $x$-direction. Additional simulations for $\lambda = \pi$ with 
$A_2 > 0.1 \lambda$ produce dense cores with sizes of the order
of the dissipation length.  The maximum density is roughly 2-3 times
higher than in the standard model (see Fig.~\ref{fig:short}) even for 
the largest wave amplitudes. This means that
fast-mode waves with wavelengths that are 2 orders of magnitude longer
than the dissipation length can only form cores in a highly turbulent
plasma. 
 
\section{Discussion and conclusions}\label{sect:discuss}
In this paper we studied the formation of density structures 
in a weakly ionised plasma by following the evolution of a fast-mode wave.
This adequately describes the formation of dense cores in a clump 
which is perturbed at its edge. Slow-mode waves propagate at low speeds
and remain near the boundary. Fast-mode waves propagate much faster
and produce slow-mode waves within a clump either due to non-linear
steepening or interactions with denser regions. However, in a weakly ionised
plasma, the ion-neutral drift is an important dissipation
process. We find that ambipolar resistivity strongly affects the evolution of 
fast-mode waves with wavelengths up to 2 orders of magnitude larger than the 
dissipation length. 

For an MHD wave in which the velocities are of the order of the 
Alfv\'en speed, the ambipolar dissipation length-scale is given by
(e.g. Osterbrock \cite{O61}; Kulsrud \& Pearce \cite{KP69}) 
\begin{equation}\label{eq:diss}
	l_d = 100\left(\frac{B}{30\mu G}\right)\left(\frac{10^{-4}}{X_i}\right)
	\left(\frac{1000}{n_H}\right)^{3/2} {\rm AU},
\end{equation}
so that $\lambda = \pi \approx$ 100~l$_d$ corresponds to 0.05 pc.
Note that this wavelength is roughly consistent with the size of a dense 
core. Fast-mode waves with $\lambda = 100 l_d$ do not produce significant 
density enhancements, even for large velocity perturbations. However,
in a sufficiently inhomogeneous medium, these waves can still produce moderately-dense 
cores with sizes of the order of the dissipation length scale. 
These cores have a short lifetime which is of the order of decades.
The existence of such short-lived, small-scale clumps has been inferred in the diffuse 
interstellar medium from the time variability of 
interstellar absorption lines (e.g. Dieter et al.~\cite{DWR76};
Crawford~\cite{C02}).

Small-scale features have also been detected in the cores of cold dense 
clouds such as in \object{Core D of TMC-1}.
Peng et al.~(\cite{P98}) identified, within TMC-1's Core D, 45 microclumps with 
sizes of 0.03-0.06~pc and masses of 0.01-0.15 M$_{\sun}$. Most of these 
microclumps have masses below the
Jeans mass. Therefore, they cannot have formed by gravitational instability.
Instead these microclumps may be generated by magnetosonic waves.

With a core density of $n_{\rm H} \approx 6\times 10^4$~cm$^{-3}$ and 
fractional ionisation $X_i \approx 10^{-6}$ 
(Caselli et al.~\cite{C98}) and $B \approx 15 \mu$G  
(Turner \& Heiles \cite{TH06}), 
$l_d = 10$~AU in Core D. The dissipation length is thus considerably smaller 
than its size of 0.1 pc, in fact waves with wavelengths up to 3 orders of 
magnitude larger than the dissipation length scale can be present in the 
core. In our models these wavelengths correspond to $\lambda = 10\pi$. 
Density perturbations are generated relatively easily by such waves, even 
for sub-Alfv\'enic velocity perturbations. This fits well with emission 
line observations of e.g. CS, NH$_3$ and CO which show a substantial 
non-thermal broadening that is not highly supersonic with respect to the 
sound speed 
in H$_2$ (e.g. Fuller \& Myers \cite{FM92}). Magnetosonic waves can thus
generate the substructure within dense cores. 

However, TMC-1's Core D has an atypical fractional ionisation. Most cores 
in the sample of Caselli et al.~(\cite{C98}) have an ionisation fraction 
which is 1 to 2 orders of magnitude lower than in Core D of TMC-1. The 
dissipation length then increases by the same order of magnitude (see 
Eq.~\ref{eq:diss}) making it unlikely that substructure is generated within 
those cores.

Lim et al.~(\cite{Limetal05}) showed that ambipolar resistivity only has 
a small, or even a negligible, influence on the evolution of fast-mode 
waves with $\lambda > 1000 l_d (= 10\pi$). The slow-mode waves 
generated by the fast-mode wave are also not affected by ambipolar diffusion
(Balsara~\cite{Balsara96}). The flow can then be described by a single 
ideal MHD fluid.  A fast-mode wave with a wavelength similar to a clump 
size ($100\pi \approx$ 5 pc) produces dense cores with 
$n_{\rm H} = 10^4-10^5$~cm$^{-3}$ and radii 
of the order 0.1 pc which are similar to observed properties (e.g.
Jijina et al.~\cite{Jijina99}). 

It is clear that, with these physical properties, most dense cores in 
our simulations have masses above the thermal Jeans mass and are, thus,
gravitationally bound. However, their masses are still lower than or of the 
same order as the magnetic Jeans mass. Magnetic pressure support then 
prevents the gravitational collapse of the core. Including self-gravity
in our models will affect the dynamical evolution of the gas, but we 
expect it to be only a local effect, i.e. the density of a dense 
core will increase. As a consequence, the dense core will become supercritical
and can undergo collapse. Simulations of the evolution of weakly ionised, 
magnetised, self-gravitating clouds justify this picture. Basu \& Ciolek 
(\cite{BC04}) and Li \& Nakamura~(\cite{LN04}) in 2D and Kudoh et 
al.~(\cite{Ketal07}) in 3D showed that sheet-like clouds fragment 
within a dynamical timescale to form supercritical dense cores surrounded 
by subcritical envelopes. However, further simulations are required to confirm  
the effect of self-gravity.

In this paper we examined the global changes to the density structure
due to ambipolar diffusion. However, our results also give a
qualitative idea of its effect on the velocity structure.
For typical clump properties the fast-mode and Alfv\'en wave with wavelengths
corresponding to dense cores dissipate their energy quickly (in 
$\approx 10^4$~yr). Slow-mode waves, however, are not subject to dissipation
(Balsara \cite{Balsara96}).The velocity 
dispersion on small scales is therefore dominated by the slow-mode wave while 
all types of waves contribute to the velocity dispersion at large scales.
By examining the eigenvectors for the slow, fast and Alfv\'en waves in 
a low-$\beta$ plasma (see e.g. Falle \& Hartquist \cite{FalleHartquist02}),
it can be shown that the velocity perturbation of a fast-mode
wave with a moderate density perturbation is of the order of 
the Alfv\'en speed, whereas it is only of the 
order of the thermal sound speed for slow-mode waves. This qualitative 
picture agrees well with the observed transition from 
supersonic motions in molecular clouds to subsonic motions
in dense cores (Myers \cite{Myers83}). 
 
\begin{acknowledgements}
We thank the anonymous referee and 
P. Caselli for useful discussions. SVL gratefully thanks 
STFC for the financial support.
\end{acknowledgements}

\end{document}